\documentclass[runningheads]{llncs}
\usepackage[T1]{fontenc}
\usepackage{graphicx}
\usepackage[export]{adjustbox}

\usepackage{booktabs}
\usepackage{times}
\usepackage{microtype}
\usepackage{hyperref}
\hypersetup{
 colorlinks=true,   
 citecolor=blue,
 urlcolor=blue,
 linkcolor=blue,
 bookmarksnumbered=true,     
 bookmarksopen=true,         
 bookmarksopenlevel=1,          
 pdfstartview=Fit,           
 pdfpagemode=UseOutlines,    
 pdfpagelayout=TwoPageRight 
}
\usepackage{multirow}

\begin{document}

\title{Knowns and Unknowns: An Experience Report on Discovering Tacit Knowledge of Maritime Surveyors}

\titlerunning{Discovering Tacit Knowledge of Maritime Surveyors}

\author{Tor Sporsem\inst{1}\orcidID{0000-0002-5230-7480}\and
Morten Hatling\inst{1}\and \\
Anastasiia Tkalich\inst{1}\orcidID{0000-0001-7391-4194}\and 
Klaas-Jan Stol\inst{1,2}}

\authorrunning{Sporsem et al.}

\institute{SINTEF Digital, 7034 Trondheim, Norway\\
\email{tor.sporsem@sintef.no}
\and University College Cork}

\maketitle

\begin{abstract}
\textbf{[Context]} 
Requirements elicitation is an essential activity to ensure that systems provide the necessary functionality to users, and that they are fit for purpose. In addition to traditional `reductionist' techniques, the use of observations and ethnography-style techniques have been proposed to identify requirements. 
\textbf{[Research Problem]} One frequently heard issue with observational techniques is that they are costly to use, as developers who would partake, would lose considerable development time. Observation also does not guarantee that all essential requirements are identified, and so luck plays a role. Very few experience reports exist to evaluate observational techniques in practice, and for organizations it is difficult to assess whether observation is a worthwhile activity, given its associated cost. 
\textbf{[Results]} This report presents experiences from DNV, a global leader providing maritime services who are renewing an information system to support its expert users. We draw on several data sources, covering insights from both developers and users. The data were collected through 9 interviews with users and developers, and over 80 hours of observation of prospective users in the maritime domain. We capture `knowns' and `unknowns' from both developers and users, and highlight the importance of observational studies.
\textbf{[Contribution]}  While observational techniques are costly to use, we conclude that essential information is uncovered, which is key for developers to understand system users and their concerns.

\end{abstract}

\section{Introduction}
\begin{quote}
    ``We can know more than we can tell''
    
    ---Polanyi \cite{polanyi2009tacit}
\end{quote}
\begin{quote}
    ``We're not good at knowing what we know''
    
    ---Ken Jennings, \textit{Jeopardy!} champion\footnote{Jennings holds the record of the longest streak of wins of the popular TV game show `Jeopardy!'} \cite{jennings2020}
\end{quote}

Organizations in all domains rely on software solutions to support their employees in achieving their core business goals. 
Automation to support professionals and experts in their manual jobs can be traced back to the early days of computing \cite{jensen1979software,tracy2021}. Initially, development processes for software were modeled after traditional engineering cycles, focusing on problem formulation and analysis, systematically developing systems based on requirements that could be identified.

Many of the early software development methods sought to provide support to systems developers in structurally and systematically design and implement systems---the so-called structured approaches; specific practices included structured analysis, structured design, data-driven design, and structured coding \cite{demarco1979,jensen1979software}. A key characteristic of these practices is that they all focus on that what can be observed and articulated; further, they take a `reductionist' approach in that these methods consider only technical entities such as components, control structures, and data flows \cite{demarco1979}, ignoring what we could label ``system-in-action'' requirements that capture the subtleties of how users actually use systems.

In recognition of the importance of capturing and managing the right system requirements, the Requirements Engineering (RE) discipline emerged \cite{mead2013history}, and has been concerned with identifying, modeling, communicating, and documenting requirements \cite{Paetsch2003}. Despite a very rich and mature RE literature today, RE remains a major challenge because systems are usually developed by people other than the intended users.
Software systems that fail to meet the needs of expert users may threaten those users' ability to do their job and, indirectly, the core business activities of the organization. A key problem is the knowledge gap that exists between the analyst/developer, and the expert user who possesses a high level of expertise. Understanding how this gap can be closed has been a longstanding goal of the RE discipline.

As one of the opening quotes suggests, Polanyi \cite{polanyi2009tacit} argued that much human knowledge acquired by highly skilled experts through experience is impossible to articulate. 
In seeking to understand whether and how expertise can be articulated, several scholars have invoked the term `tacit knowledge' \cite{gacitua2009making}. 
Gervasi et al. \cite{Gervasi2013} drew on a notable 2002 press briefing of the late Donald Rumsfeld, Secretary of Defense during the U.S. invasion of Afghanistan and Iraq. Rumsfeld argued there are different types of knowledge, or `knowns': known knowns, known unknowns, and unknown unknowns.\footnote{These different types of `knowns' map very well to Phillip Armour's ``Orders of Ignorance'' published two years prior, in 2000 \cite{armour2000five}. This might be a rare unintended instance where SE research has had an impact on global political rhetoric.} Gervasi et al. \cite{Gervasi2013} suggested that there is a fourth type of knowledge: 
\begin{quote}
\textit{``An unknown known is knowledge that a customer holds but which they withhold from the analyst.''}
\end{quote}
We note that this `withholding' is likely to be unintentional. Tacit knowledge, then, fits that definition, i.e. tacit knowledge is an unknown known \cite{Gervasi2013}. Table~\ref{tab:knowns} presents the four types of knowns, considering two important roles in RE, as Gervasi et al. \cite{Gervasi2013} suggested: the system analyst/developer, and the  user. Some knowledge is held by both developers and users (known knowns), whereas other knowledge is known to only one but not the other, e.g. unknown knowns represent knowledge held by users, whether they are aware of it or not, but unknown to developers.

\begin{table}[!th]
    \caption{Developers and users' knowns and unknowns (based on Gervasi et al. \cite{Gervasi2013} and Sutcliffe and Sawyer \cite{sutcliffe2013requirements})}
    \label{tab:knowns}
    \begin{tabular}{p{0.7cm}p{1.3cm}p{4.2cm}p{4.2cm}}
    \toprule 
    & & \multicolumn{2}{c}{Analysts and Developers}\\
    \cmidrule{3-4} 
    & &  \multicolumn{1}{c}{Known to developers} & \multicolumn{1}{c}{Unknown to developers}\\
    \cmidrule{2-4}
    \multirow{10}{=}{Users}     & Known to users & \textbf{Known Knowns}: relevant knowledge that users know  and that can be articulated for software developers & \textbf{Unknown Knowns}: relevant knowledge that users know (whether consciously or without realizing it), but which is not yet articulated and thus not known yet to software developers.\\
    \cmidrule{2-4}
    & Unknown to users & \textbf{Known Unknowns}: relevant information that developers are aware of (know), but which they don't know yet. Users may be unaware of this knowledge, or have forgotten it. & \textbf{Unknown Unknowns}: potentially relevant information, but both developer and user are unaware that it is missing. Developers lack relevant domain knowledge, and users are unaware of the knowledge that they rely on.\\
    \bottomrule 
    \end{tabular}

\end{table}

The RE field has discussed different requirements elicitation techniques at length \cite{davis2006effectiveness,hickey2003elicitation,sutcliffe2013requirements} and it lies beyond the scope of this paper to present a full discussion.
Commonly discussed techniques are interviews, workshops, scenarios, and observation \cite{davis2006effectiveness,hickey2003elicitation,sutcliffe2013requirements,zachos2005rich}. Observation is often mentioned as a part of conducting ethnographic studies; several papers have discussed ethnography or observational approaches to support requirements elicitation and design \cite{hughes1995presenting,hughes1992faltering,sommerville1993integrating}.
Early studies proposing to integrate ethnography for RE recognized that traditional techniques \textit{``do not take into account actual work practices''}  \cite{sommerville1993integrating}. 
While there has been some fruitful discussion and analysis of observational methods to uncover `unknowns,' a few issues seem to remain. For example, ethnographically-informed or observational methods for RE have been suggested to require considerable resources, time in particular, making them less attractive. Further, other issues  associated with ethnographic research is that it may suffer from ambiguous interpretation \cite{sutcliffe2013requirements}, a lack of technical competence of ethnographers \cite{crabtree2002}, and the serendipitous nature of identifying new requirements through ethnography \cite{sutcliffe2013requirements}, i.e. the reliance on luck. Finally, the number of studies that evaluate the use of observational approaches including ethnography for requirements engineering has remained limited, despite several important contributions in the 1990s and early 2000s \cite{hughes1992faltering,hughes1995ethnography,hughes1995presenting,crabtree2002}. One might wonder, given the drawbacks listed above, whether organizations should bother with observational approaches. A lack of experience reports on the use of observational or ethnographic techniques hinders organizations in deciding whether this approach is worth the considerable cost. There seems to be an acceptance that there is no advantage in any specific technique over the use of structured interviews \cite{hickey2003elicitation,sutcliffe2013requirements}, and so an open question is: what value does observation offer in a requirements engineering context? 

Thus, the goal of this experience report is to highlight the importance of observation to identify unknown knowns and unknowns, and report lessons learned from the field, in a domain that hitherto has not been studied in this context. Several of the seminal papers in the software engineering and requirements engineering literature reported on an air traffic control system, which represents a very specific setting whereby its users operate in a fixed location. This experience report focuses on surveyors who inspect ships for certification, necessary to allow them to operate in international waters. Surveyors, unlike air traffic controllers, operate in a different setting \textit{every single day}. This makes characterizing these actors' work environment more challenging as each ship is unique and thus it is important to recognize the varying work settings of these experts. We illustrate how  developers, who had used traditional requirements elicitation techniques such as interviews, were struggling to understand these expert users, and indeed had not gained important insights that we classified as unknown knowns and unknown unknowns (see Table~\ref{tab:knowns}). We conclude by juxtaposing our findings with prior literature, adding clarifications and commentary, and identify some implications for practice.

\section{Methods}
This experience report draws on data collected from different sources at DNV, a major service provider in the maritime sector. As researchers of the SINTEF Digital Process Innovation group, we are involved in an ongoing project with DNV focused on Digital Transformation, which provided the backdrop of this investigation. 
In the remainder of this section we describe DNV and procedures for data collection and analysis.

\subsection{Description of DNV}
DNV  is a leading service provider in the maritime sector, with about 3,700 employees operating globally. DNV's core business is compliance verification of vessels (ships of any size); successful verification leads to issuing of necessary certificates that vessels require in order to secure marine insurance and sail and operate in international waters.  
Certificates are normally issued annually, with a more thorough five-year survey. Vessels are costly to run; therefore, they are continuously in operation. Surveys are typically conducted during visits to ports or shipyards, when vessels load or unload cargo, or undergo maintenance. Every survey job is tailored to the unique characteristics of a vessel, and its operation plan in order to reduce the interruption to normal operations. This means that survey procedures are often broken down into parts, with each part of the survey potentially being conducted in a different port.

DNV is currently modernizing its survey support system, which surveyors use to conduct and manage surveys. The system is used for planning survey jobs, document compliance, reporting of `findings,' (that is, issues that require fixing before compliance can be signed off), looking up a vessel's history, and issuing of certificates. 
The current desktop version for Microsoft Windows was released in 2004, and at the time of data collection, DNV was developing a new web-based solution to  allow continuous development of new features. Development is organized as an in-house project with a release date when the new solution goes live and the old system shuts down. 
DNV employs approximately 1,000 surveyors globally who are the primary users. 
This group of users tend to dislike new digital tools -- or in the words of one surveyor, \textit{``we don't like change.''} 
DNV management was concerned that if the new system gained a bad reputation, the cost would rise dramatically, possibly outweighing the benefits of the new system, requiring significant resources to overcome  resistance in adoption. In other words, management put a premium on developing a system that pleases its intended users.

\subsection{Data Collection and Analysis Procedures}

We collected data during a nine-month period; data collection and analysis were interleaved, and followed procedures described by Seaman \cite{seaman1999qualitative}.
The data collection activities included semi-structured interviews and several site visits for observation of surveyors at work. The first site visit for observation was treated as a pilot study, and from this we gained valuable insights into how this group of users interacts with software technology and a general understanding of their role; based on this we designed observation guides and semi-structured interview guides. 

Interviews can be a valuable source of information as it allows in-depth conversation with experts, but it is only one of many potential methods to collect data in field studies \cite{lethbridge2005studying}.
Interviews fall in a category of methods that Lethbridge et al. have labeled \textit{inquisitive} techniques \cite{lethbridge2005studying}, in that a researcher must actively engage with interviewees to get information from them. A second category is \textit{observational} techniques, which includes  observation of professionals. Both types of techniques have benefits and drawbacks; interview data may be less accurate than observational data, but observational techniques may introduce the Hawthorne effect, whereby professionals' processes change when they are observed \cite{lethbridge2005studying}.

A total of nine interviews were conducted: four software developers, one manager, one implementation manager, and three surveyors. The focus of these interviews was to develop an understanding of the purpose of the new system and how developers elicited user requirements. All interviews were transcribed, resulting in 105 
pages of text. 
Following the interviews, we conducted a total of seven observations. We observed three more surveyors for two days each, and three surveyors for one day each. Three of the surveyors were situated in Norway and four in the Netherlands.

The first two authors conducted observations of seven surveyors. These onsite activities were organized in collaboration with DNV's central scheduler who assigns jobs to surveyors. 
The site visits involved shadowing the surveyors for the full day; this included accompanying surveyors during inspection of vessels, including
crawling through narrow storage tanks, climbing crane towers, as well as driving for hours to reach remote ports during which surveyors could also have phone calls with colleagues, and having lunch together.
Our impression is that the surveyors appreciated the opportunity to show their work practices and expressed themselves freely. 
The researchers conducting observations were dressed similarly to crew and surveyors, including all the required Personal Protective Equipment (PPE) (including safety helmet, ear muffs, safety shoes, etc.). 
In a way, we were more like apprentices than researchers. 
Research notes and pictures were constantly captured over the course of data collection, and reflections were written immediately afterwards, resulting in 59 pages of 
notes produced and about 100 photographs of surveyors in action (see Figure~\ref{fig2}).

The first and second authors jointly analyzed the data and immersed themselves in the material. A word processor  was used for both open-ended coding and memoing \cite{seaman1999qualitative}. Examples of labels include:
\begin{itemize}
    \item ``use of phone calls, not chat, to maximize bandwidth of communication and realtime feedback''
    \item ``surroundings force surveyors to take breaks during their work day''
\end{itemize}
These two labels were grouped in a theme ``adapting to surroundings.''  
After we completed the data analysis, we used member checking, a procedure to assess the validity of our findings by presenting them 
in a workshop involving surveyors and DNV management, and adjust any misapprehensions. Overall, their response was confirmative.

\begingroup
\setlength{\tabcolsep}{2pt}
\begin{figure}[!t]

\begin{tabular}{@{}lr@{}}
\includegraphics[height=5.15cm]{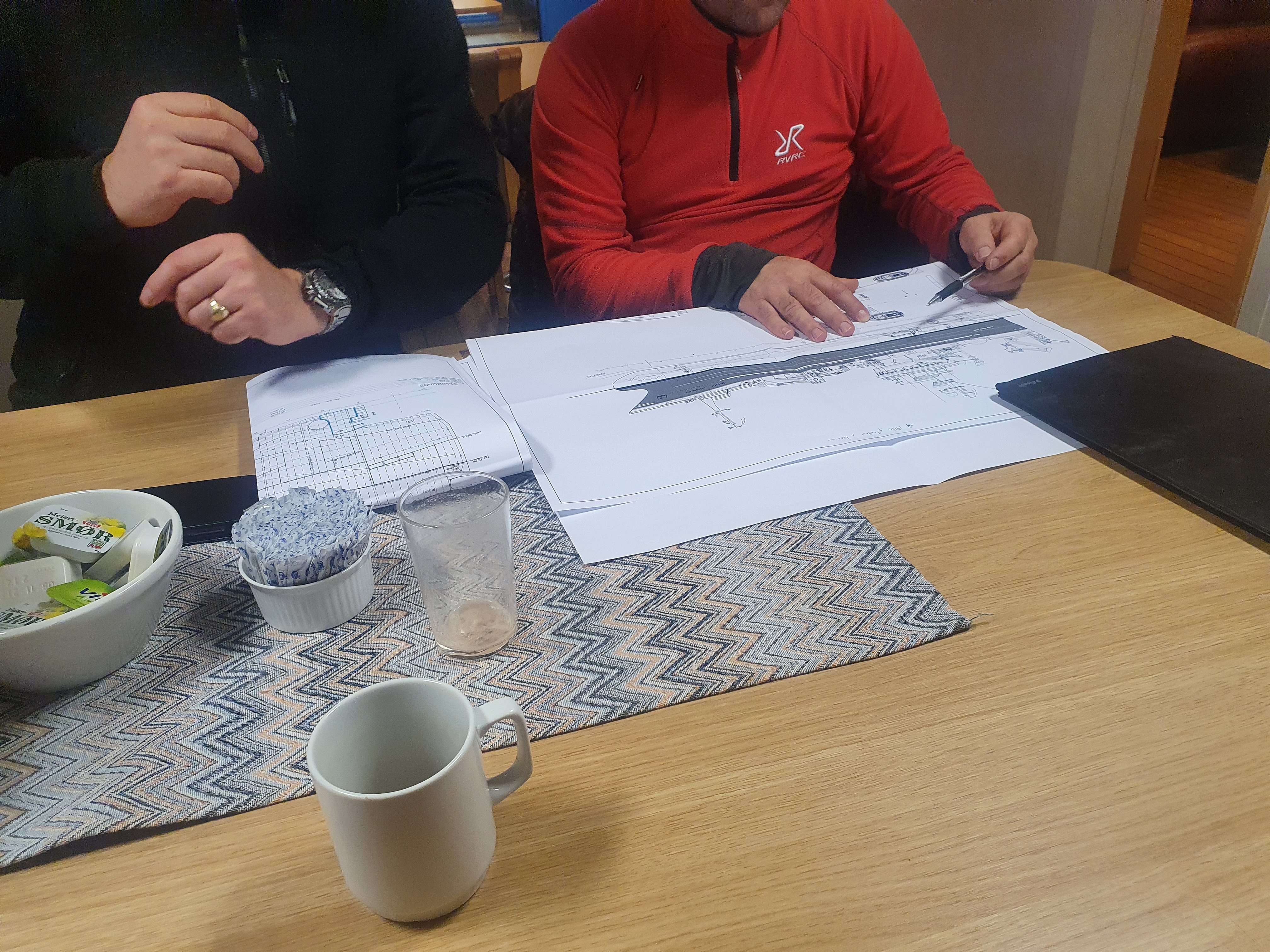} &
\includegraphics[height=5.15cm]{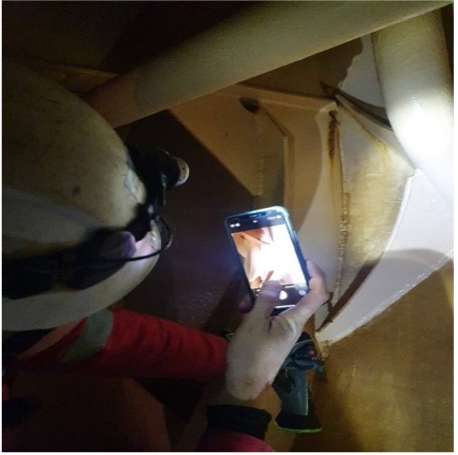} \\
\end{tabular}

\begin{tabular}{@{}lcr@{}}
\includegraphics[height=4.13cm]{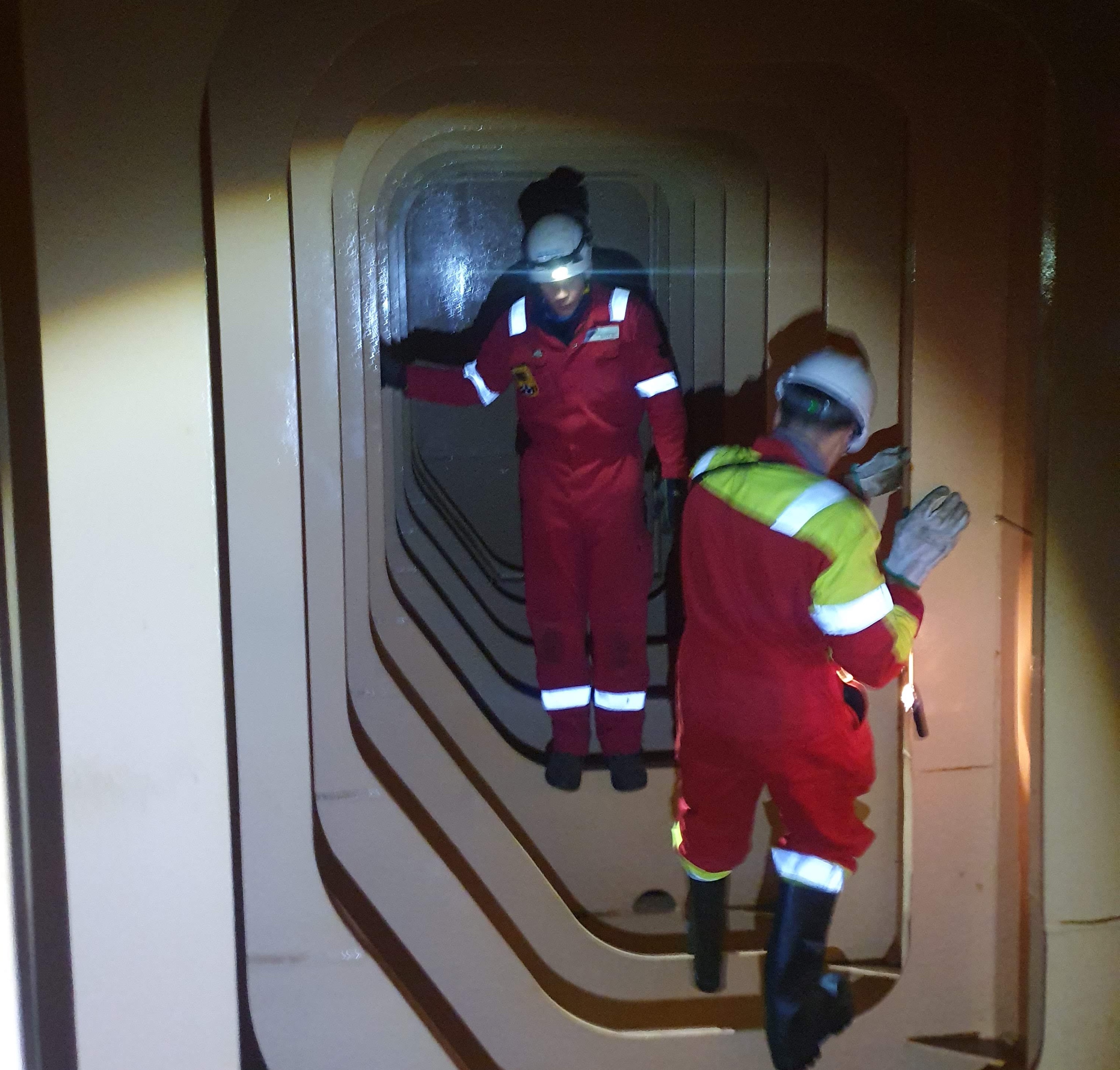} &
\includegraphics[height=4.13cm]{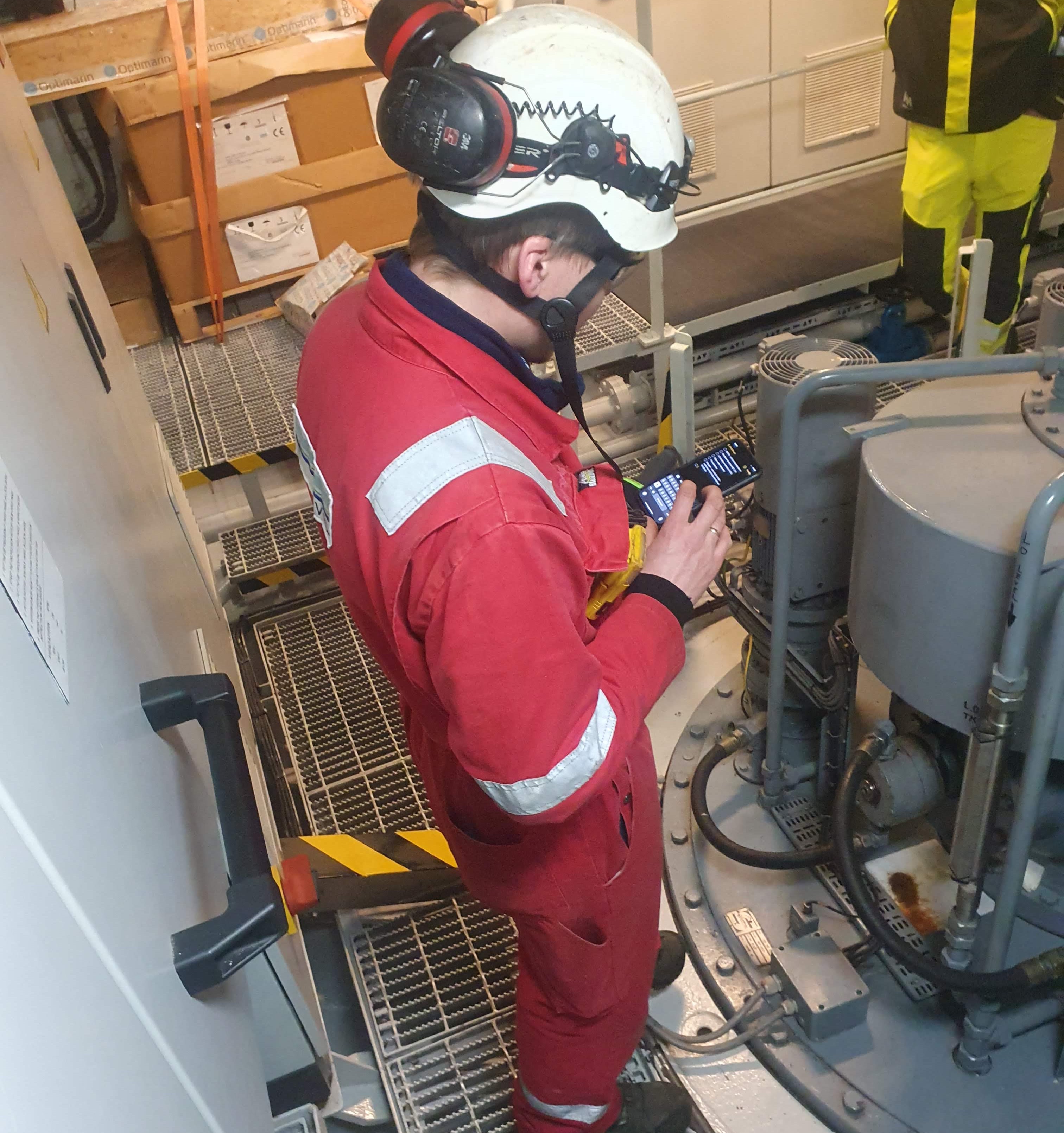} &
\includegraphics[height=4.13cm]{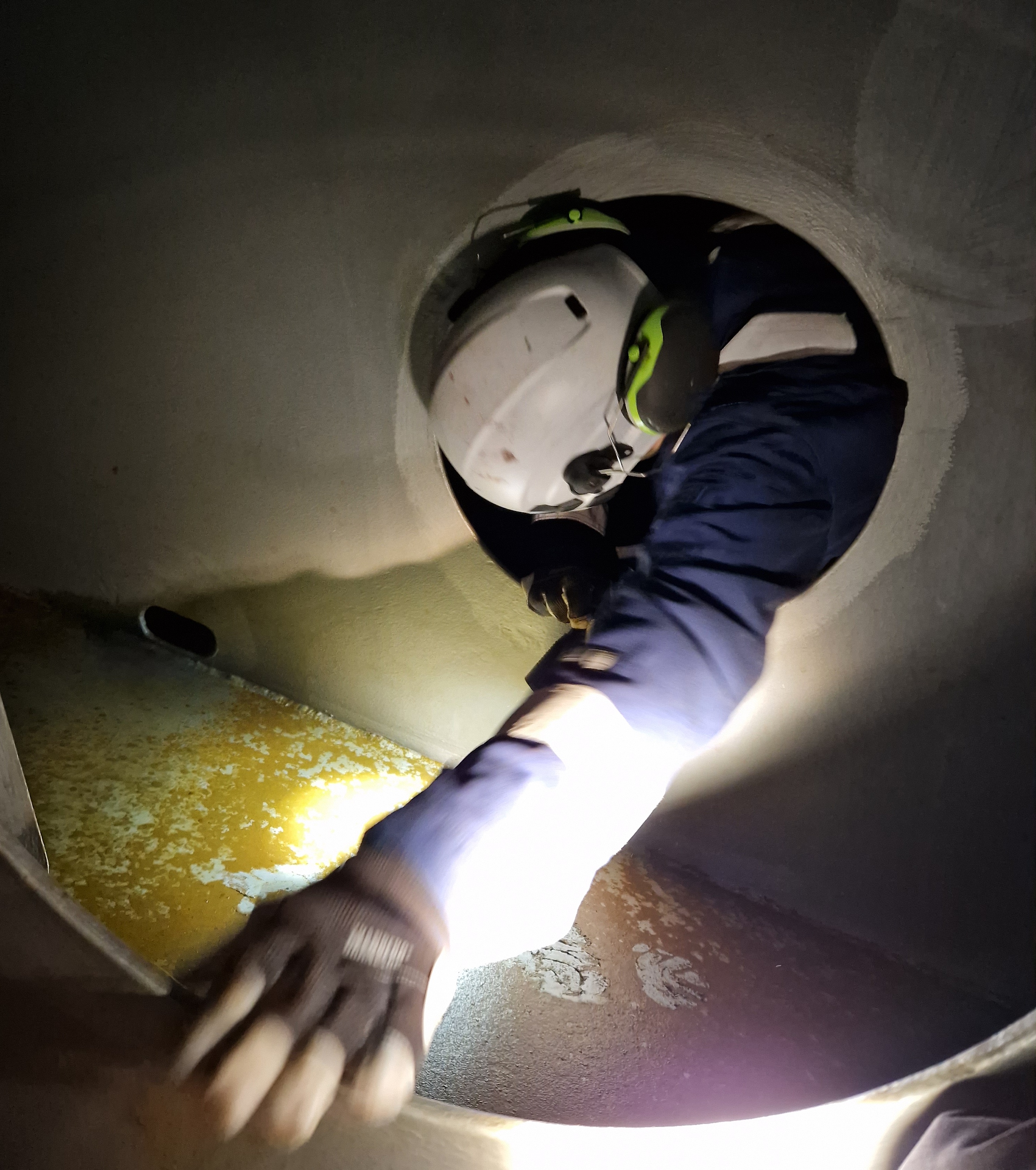}\\
\end{tabular}

\begin{tabular}{@{}lcr@{}}
\includegraphics[height=4.176cm]{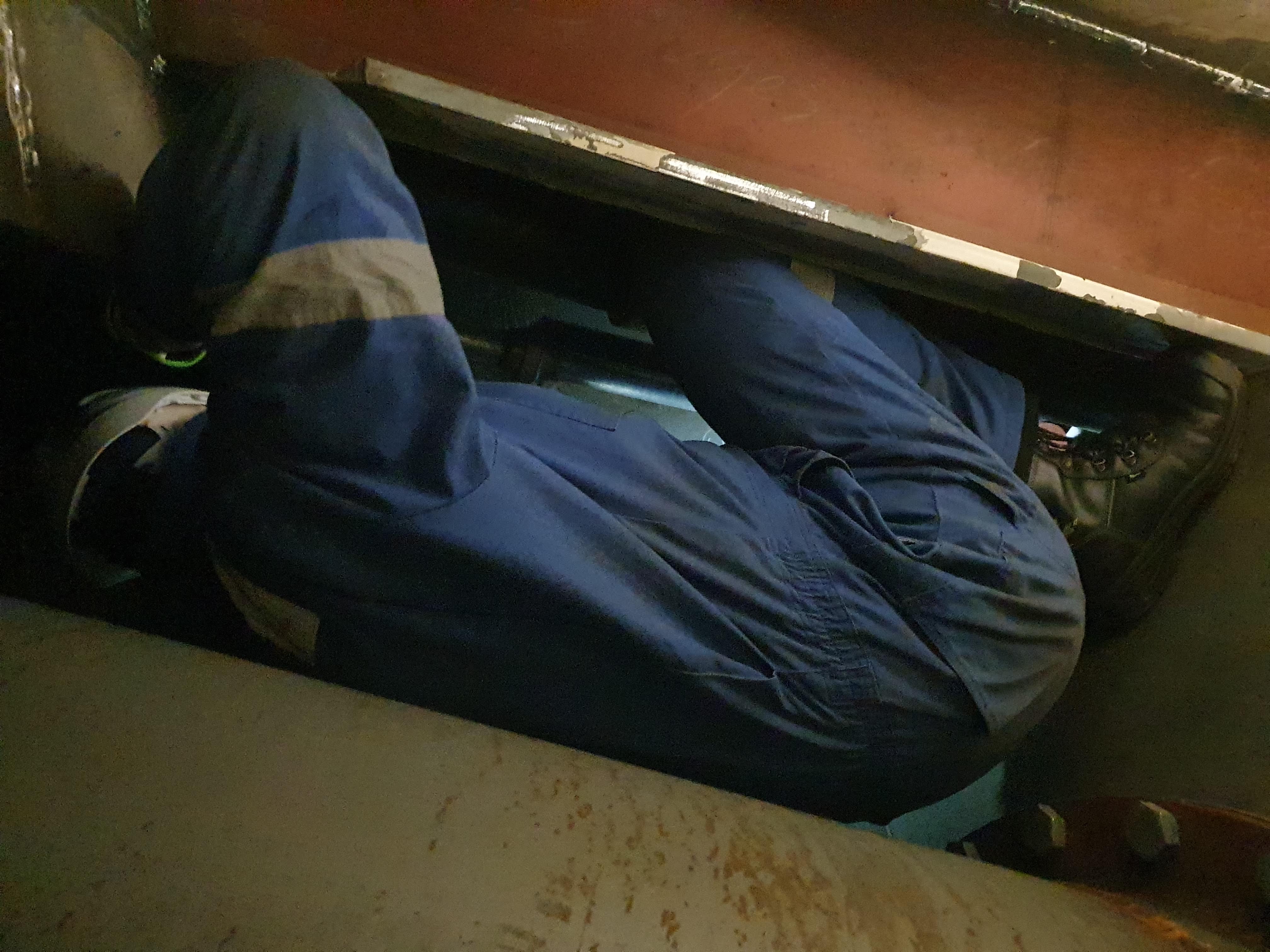} &
\includegraphics[height=4.176cm]{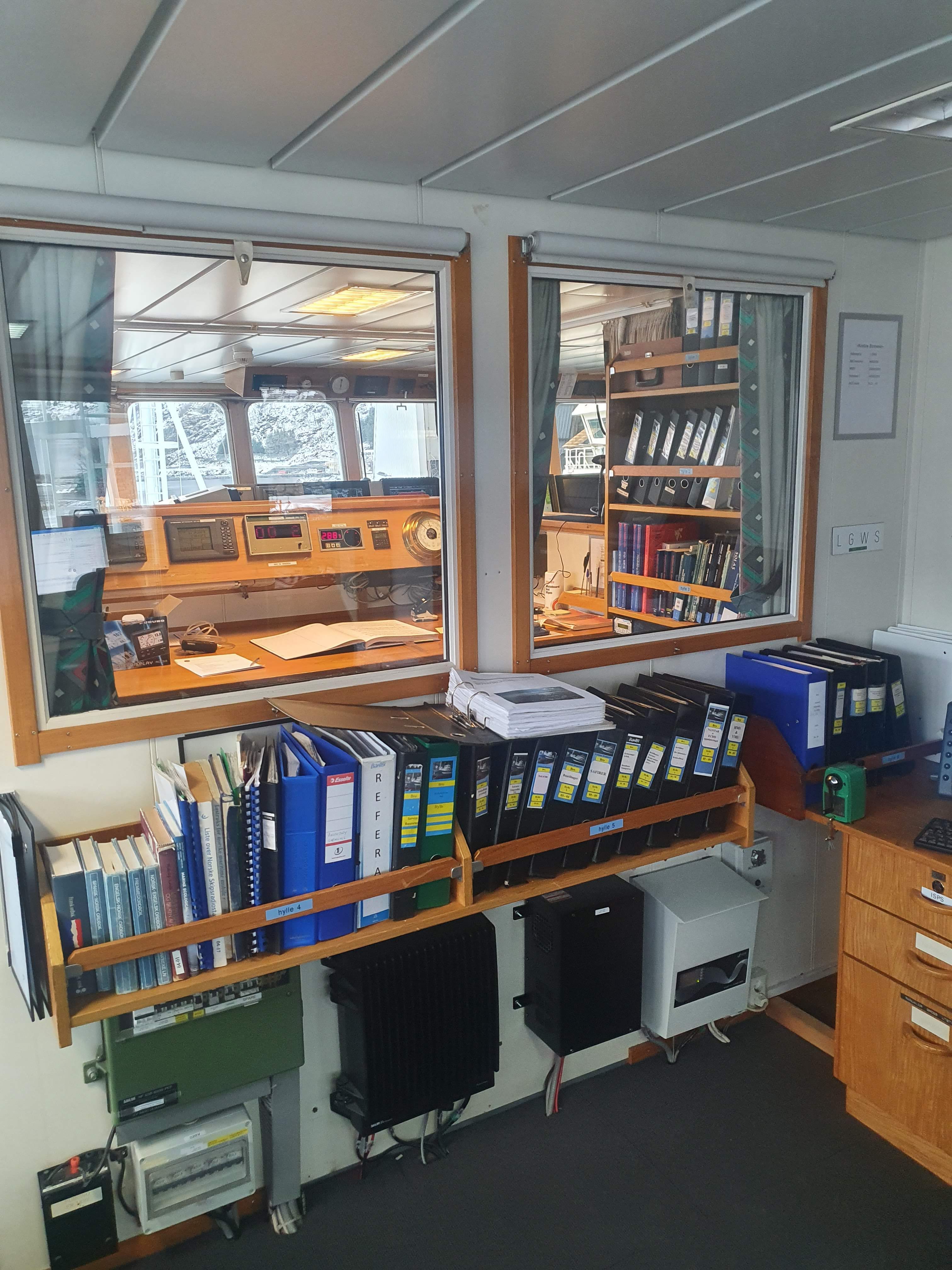} & 
\includegraphics[height=4.176cm]{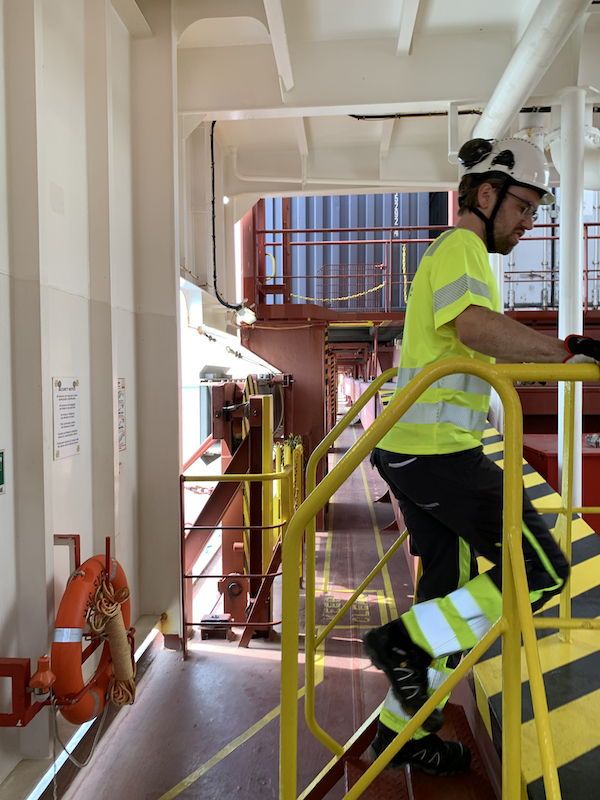}\\
\end{tabular}

\caption{From left to right: planning the survey, photographing issues, surveying in difficult environments, paper artifacts remain important, and one of the researchers on-site. No photographs could be taken on tankers as we had no explosive-proof camera.} \label{fig2}
\end{figure}

\endgroup 

\section{Findings}
This section presents the key findings. We first discuss requirements elicitation practices at DNV; the remainder of the section is organized using the framework presented in Table~\ref{tab:knowns}; a summary of findings is presented in Table~\ref{tab:findings2}.

It should be clear that the `users' in this context are domain experts, namely, surveyors of the DNV organization, who have very extensive experience; the term `users' does not therefore apply to other types of users.

We do not discuss Known Knowns in further detail, because that is knowledge shared among developers and surveyors alike. One example of this is that surveyors preferred to keep the new system as similar to the old one, and developers were aware of this and tried to accommodate this.

\begin{table}[!t]
    \caption{Summary of Findings}
    \label{tab:findings2}
    \begin{tabular}{p{1.2cm}p{4.5cm}p{5.3cm}}
    \toprule 
    &  \multicolumn{1}{c}{Known to developers} & \multicolumn{1}{c}{Unknown to developers}\\
    \cmidrule{1-3}
    Known to users 
    & \textbf{Known Knowns}:
    \begin{itemize}
        \item Surveyors preferred to keep the new software as similar as possible to the old one, which developers knew and tried to accommodate. (Not discussed in the findings section, because this is unproblematic knowledge.)
    \end{itemize}
    
    & \textbf{Unknown Knowns}: 
    \begin{itemize}
        
    \item Surveyors rely on ``gut feeling'' and ``on the go'' decision making.
    
    \item Surveyors prefer to discuss issues over the phone with colleagues, instead of in writing (e.g. chat or email), ignoring the `chat' function that was designed for this.
    
    \item Surveyors spend much time interacting with people onboard of vessels, which is essential for a successful survey, but is hard if not impossible to capture.
    \end{itemize}\\
    
    \cmidrule{1-3}
    Unknown to users 
    & \textbf{Known Unknowns}: 
    \begin{itemize}
        \item Developers frustrated with an inability to test their assumptions.
        \item Developers lacked domain knowledge and realized they were missing information when merely talking to surveyors.
        \item Developers envisioned the user according to how they understood them from the workshop and interviews, filling in blanks using their own logic rather than by asking the surveyor, and their experience might be limited.
    \end{itemize}
    & \textbf{Unknown Unknowns}: 
    \begin{itemize}
        \item Divergence between observational and interview data: surveyors simplified, generalized, and abstracted when talking about their job, but reality is different.
    
        \item Surveyors must constantly adapt their workflow to changing circumstances; they survey process is not straightforward and is tailored to the context.
        
        \item The application to report findings (issues that require fixing) technically works, but is not used as originally conceived by designers due to inconvenient menu navigation and illogical object naming.
    \end{itemize}\\
    \bottomrule 
    \end{tabular}

\end{table}

\subsection{Requirements Elicitation at DNV}

The team we interacted with did not have any dedicated requirement engineers; requirements therefore were elicited by the developers.
Development teams relied mainly on two methods to capture the surveyors' requirements.

\begin{enumerate}
    \item  \textbf{Workshop.} A three-day workshop with a user representatives group of 50 surveyors face to face in the early phases to gather as much information as possible about how surveyors work and interact with technology.
    One developer explained their focus on the gap between current features and what surveyors need and expect: \textit{``We were discussing current solutions and what they [the surveyors] miss.''} 
    Following the workshop, a UX-designer created user stories based on the results. 
    
    \item \textbf{User tests.}	Variants of one-on-one test sessions virtually on Microsoft Teams to test usability and functionality developed from the user stories gathered in the workshop. One surveyor in the user representative group explained:
    
    \begin{quote}
        \textit{``I share my screen and they sit and take notes along the way or ask [questions]. We did tests where I kind of got instructions on what to do (...) and tests to check if I intuitively could find what to do.''}
    \end{quote}
    
    These tests were facilitated by the UX-designer and ranged from strictly orchestrated, to tests where the user was given a task and encouraged to explore the system on their own to solve it. 
    Tests were conducted every 2-3 months with the same selection of surveyors as the one attending the first workshop.

\end{enumerate}

\subsection{Known Unknowns: What Developers Know They Don't Know}
Developers were aware that they did not know certain aspects. To better develop and gain insights into these known gaps in their knowledge (`Known unknowns') developers had continued access to the user representative group during development. This proved useful when developers sought to ask for clarifications while reading user stories and developing features. 
One of the developers explained that: 
\textit{``Some assumptions that we've had before  proved slightly different.''} 

Developers described this combination of observation during user tests on the one hand,  and the ability to contact surveyors directly for follow-up questions on the other hand, as `ground-breaking' because they had not had such close contact with users before. 
At the same time, they sought to acquire an even deeper insight into the user's context and observe surveyors use their software in the real world, to understand \textit{``How does this [software] relate to how they actually work?''}  
Developers were aware they lacked this understanding, and expressed a preference to visit the world of surveyors and conduct observations themselves. One developer shared that: \textit{``I have not been on a tanker before, so I have no idea what things really look like.''} 

Developers argued they could not obtain all the essential information about user needs due to a lack of basic domain knowledge of surveying. One of the developers recognized that context information was crucial to gaining a deeper understanding of the survey process in practice: 
\begin{quote}
    \textit{``I would like to get on a boat and see what the actual work process looks like [...] there is a lot they [surveyors] cannot include when they explain it in the office, versus when you are actually out physically with them.''}
\end{quote}

At the same time, observing a surveyor in real life is not without challenges, not least of which include the cost associated with site visits as well as the purchasing of prerequisite PPE, and the cost associated with lost developer productivity for the time they travel. 
One developer highlighted: 
\begin{quote}
    \textit{``We have always requested that we visit a ship so that we can actually connect the dots between our domain knowledge versus what we actually see in practice.''} 
\end{quote}
Developers were frustrated that some of their work had to rely on assumptions that were impossible to test through traditional methods such as workshops and interviews. It seemed there was a shared recognition among the developers that if they could not test these assumptions, the software would not fulfill its potential.

\subsection{Unknown Knowns: What Developers Don't Know, But Users Do}
There was also a category of knowledge that developers were not aware of, but the surveyors were. 
All surveyors we shadowed were highly experienced in their role. When boarding a vessel, they would quickly gain a `feeling' of the vessel's condition, as they had learned to recognize subtle clues of technical problems (``findings'') through observations and conversations after years of experience. Subtle clues include the freshness of the paint, the general tidiness of the deck, the condition of the lights, and signs of stress among the vessel's captain and crew.
By piecing together these clues, surveyors  adapted the survey job to uncover the most crucial findings. For example, one of the surveyors we followed decided to check all so-called ex-lights (lights certified for explosive environments) after having been only minutes on board a vessel. The surveyor then continued to make numerous other findings. When we asked why he decided to go straight for the lights, he said that it was a combination of experience and pieces of information he gathered when boarding the vessel, concluding: \textit{``I get this gut feeling.''} 
He did not need to spend time analyzing what to focus on in his survey but intuitively made this decision \textit{``on the go''}---the checklist or survey support system did not prompt the surveyor.
This was knowledge he found impossible to describe to developers, arguing: \textit{``You can only learn it through years of experience.''}
We made similar observations with other surveyors, all of whom shared the same explanation. 

Although surveyors tried to explain this type of knowledge during interviews, it was not until we observed this ourselves, while attending surveys on vessels, that we gained an adequate understanding of how such knowledge impacted the way the survey was conducted. It had remained unknown to us as researchers and to the developers. 

Because the surveyors struggled to communicate many critical aspects of their work to others (non-surveyors), they felt they were unable to articulate their needs clearly to developers. 
Previously, DNV had digitalized the old paper-based checklists to relieve surveyors of hours at the printer. Despite knowing most checklists by heart and rarely printing them, surveyors welcomed this improved accessibility. However, whereas the old checklists automatically marked newly added check items in red, this feature disappeared with the introduction of a new system; this forced surveyors to search for this information that was now `hidden' within hierarchies of application menus. 
Surveyors agreed that a lack of understanding of surveyor work among software developers was the reason for this shortcoming.

Another crucial part of surveying, that remained completely unknown to developers, is how surveyors establish relations with the captain and crew upon entering a vessel. This is important because both captain and crew act as `gatekeepers' as well as facilitators for the survey. Good working relations results in a smoother survey because the captain and crew willingly support the surveyor in accessing the vessel's different parts, i.e. by opening hatches, stopping maintenance work, clearing gas tanks, etc. They are powerful allies because they can provide flexibility to the surveyor. Surveyors and crew constantly negotiate which survey activities are convenient based on the current situation on the vessel. For instance, when a surveyor planned to inspect tanks, a cooperative relationship with the Chief Officer gained him increased access and guidance during the survey.

Establishing good relations happens in social situations. Typically, when surveyors board a vessel they go straight to the bridge and meet the captain and First Officer. Coffee is offered, sometimes cakes, and usually a polite conversation ensues about for example the vessel's history and mutual acquaintances. They then move to planning the day and negotiating what surveyor activities are possible considering the vessel's operations that day. Drawings of the vessel and survey-checklists are commonly central artifacts for achieving a shared vision amongst them, because they are able to show and tell, pointing to details as a basis for discussions (see Fig.~\ref{fig2}). These artefacts are found either on the surveyor's computer screen or printed by the captain to make it easier for them to gather around.

Social situations, like these, remain unknown to developers because surveyors perceive this to be an informal part of the survey process, and as irrelevant to developers: 
\begin{quote}
    \textit{``I spend half my time going around the vessel talking to people. That human part is not very well captured [in work instructions, procedures, etc.].''} 
\end{quote}
This becomes a challenge because developers do not know that the software they develop is a critical artefact in establishing good relations between surveyors and captains. 

Another example of unknown social interactions occurs when surveyors talk on the phone. Surveyors are dealing with problem solving in complex surroundings and often need to discuss their situation with colleague surveyors. In a time-pressed world they reach out for colleagues because they are able to explain their problem and surroundings within minutes, let their colleagues ask questions, and interpret the situation together to agree on how to proceed. Such conversations contain valuable and almost-instant information about problem-solving that DNV attempted to capture by introducing a chat function. By capturing these conversations through chats in the application, knowledge could be saved and processed to benefit other surveyors experiencing similar problems. However, surveyors only used the chat for straightforward and simple issues, leaving the more complicated issues for phone conversations, meaning the most valuable and interesting conversations were never captured. 
One surveyor argued that:
\begin{quote}
\textit{``It [a phone call] creates fewer misunderstandings, less dissatisfaction, more understanding and makes it easier to reach an agreement. You don't waste time writing e-mails''} 
\end{quote}
We observed such conversations between surveyors and actors like captains, engineers, managers, superintendents, ship owners, authorities, and colleague surveyors. Characteristics of such phone calls remained unknown and indeed invisible to developers.
 
These were some examples of critical pieces of expert knowledge about surveyors' daily jobs, constraints and requirements that developers simply were not aware of, and that surveyors struggled to articulate during requirements elicitation.

\subsection{Unknown Unknowns: What Neither Developers Nor Users Know}
Finally, there was a category of knowledge that neither developers nor the surveyors were aware of, the unknown unknowns. 
The surveyors' workflow was highly dependent on, and tailored to a complex context with constantly changing circumstances onboard the vessels. This was an important issue for developers to understand, yet they were not aware of this, and nor were surveyors readily aware of this as they described their work in interviews.
For example, surveyors were forced to take short spontaneous breaks caused by changing circumstances on deck. On one occasion, they had to wait for a crew performing gas measurements before the surveyor could enter a tank---strict protocols are in place before people may enter certain  hazardous areas of a vessel. The loading and unloading of cargo also affected the survey activities that could be performed; bunkering (refuelling of a vessel) could also limit certain activities, such as a black-out test that tests back-up generators.
Surveyors would use these breaks to capture findings. Although  surveyors were expected to use the application to do this, we observed that this involved large numbers of clicks and taps when navigating menus to register findings. Surveyors found this too cumbersome, because breaks were usually short and unpredictable. In addition, several surveyors  found it challenging to locate the correct `object' in the application because their naming did not always make sense to them. Therefore, they avoided using the application during inspections and instead preferred to register their findings manually, for example by taking photographs of issues, or handwritten notes.

Although the application seemed to be an excellent digitalization effort in theory, it failed to support the experts in practice because it was incompatible with how they performed their job in a complex and unpredictable environment. The surveyors perceived that the system failed because their context was too complex to describe all of its aspects to software developers. 
Interestingly, when we compared our observational data and interview data, we discovered significant divergences in how surveyors \textit{said} they worked and how they \textit{actually} worked. 
During interviews, surveyors would generalize and simplify how they worked, leaving out details and abstracting away aspects that are hard to comprehend without field experience. During observation, however, they referred to situations they participated in as a starting point for more detailed explanations.  They were not able to translate their expert needs to software requirements. They acknowledged that these requirements remained unknown to them and that developers would benefit by making observations in the real world. 
One of the surveyors proposed that \textit{``They [developers] have to come out here and see for themselves.''}

In sum, both developers and surveyors recognized the need for developers to observe the world of surveyors to better understand the needs of this group of expert users.

\section{Discussion and Conclusion}

\begin{table}[!tb]
    \caption{Contributions and Implications}
    \label{tab:implications}
    \begin{tabular}{p{3.6cm}p{3.8cm}p{3.5cm}}
    \toprule 
    Insights from prior literature & Findings of this Study & Implications for Practice\\
    \midrule 
    
    Observation and ethnography are resource-intensive activities. Early scholars proposed `quick and dirty' approaches'' \cite{hughes1995presenting}. & The gap between developers and users may be too large to bridge; observing surveyors even for a few days provided valuable insights that are difficult to articulate. & The cost associated with letting developers go into the field for even a week may well be worth it, potentially improving the quality of requirements analysis.\\
    
    \addlinespace
    
    Observation and ethnography depend on `luck' and serendipity to identify requirements \cite{sutcliffe2013requirements}. 
    & Observing expert surveyors at work for even a few days provided deep insights into the constraints and work practices. & Observation is not only about identifying requirements and features, rather, it can be valuable for developers to understand the real-world context and constraints (see Fig.~\ref{fig2}). \\
    
    \addlinespace 
    
    Integrating ethnographic observations into structured methods of requirements analysis is very challenging \cite{sommerville1993integrating}. The reductionist character of RE (focusing on components, data flows, processes) is not compatible with ethnographic inquiry \cite{crabtree2002}. &
    Surveyors prefer ad hoc communication \textit{outside} the system's functionality (rather than built-in features e.g. chat). The social interactions and goodwill between surveyor and captain/crew are essential for a successful survey, which cannot be captured in system features. 
    & Rather than seeking ways to integrate observational findings into requirements analysis, consider system boundaries and develop a good understanding of the social context where systems are implemented. \\
    
    \addlinespace
    Experts find it difficult to articulate their expertise during requirements analysis \cite{sommerville1993integrating}. Ethnography recognizes work activities as they are actually conducted, rather than some idealized version of it \cite{hughes1995ethnography}. & Surveyors frequently rely on `gut' instincts developed over years of experience. Surveyors are often assigned jobs on short notice, and have to adjust surveys based on continuously changing circumstances.  & Not all work practices can be digitalized. The valuable experience and gut instincts cannot be replaced with a rigid workflow. Develop systems that empower experts, rather than aiming to digitalize existing workflows. \\
    
    \bottomrule 
    \end{tabular}

\end{table}

Most prior work in the software engineering domain was conducted by a relatively limited number of authors, detailed many insights from a select number of case studies, in particular studies of air traffic controllers \cite{hughes1995ethnography}. Several other studies are situated in the HCI and CSCW communities in the 1990s and early 2000s, recording lessons for designing interactive systems that became increasingly popular in the 1990s \cite{coad1990}.

In this paper we report a number of insights (see Table~\ref{tab:implications}). For example, whereas prior literature has suggested the high cost of ethnographic or observational studies might preclude organizations from using these strategies, we would argue that in some domains the gap between developers and users is simply too large to bridge. Even a few days of observing surveyors at work provided extensive insights that could have saved many hours of developer time. It is important at this point to distinguish between `observation' and `ethnography'; the former being a technique that we focused on primarily, whereas the latter is a more encompassing strategy that would take far more time. 

A second issue with observation is that it depends on serendipity and `lucky'  circumstances that would lead to identifying new requirements. While we agree this is an issue in \textit{identifying} requirements, it is less of an issue to \textit{understanding} the nature, context, and constraints of an expert's daily job. We argue this is an important distinction to keep in mind when planning site visits for developers or requirement engineers; the goal of observation then becomes one of ``walking in the user's shoes,'' to understand the system-in-action context before attempting to capture requirements.

Analysts and developers are trained professionals who look at systems through a lens of the \textit{affordances} that technologies offer, but this too has a reductionist `smell': looking at technologies and how they ``map'' to possibilities, rather than user preferences. It is possible, of course, to add a chat function that seeks to capture potentially valuable conversations---but this ignores users' \textit{preference} to \textit{not} use system features, but use simple means such as a phone for a direct and possibly private conversation. 

Finally, experts are known to experience difficulty articulating their knowledge, certainly when this includes dependency on their gut instincts that developed over years. No guidelines, handbook, or IT system can replace this. This leads us to argue that, perhaps, not all work practices can (and should) be digitalized. In a way, we are touching upon the boundary of a current popular trend in the IT industry, namely a search to digitalize everything. While we do not deny that software systems can greatly improve our lives and productivity, care should be taken to understand how software solutions can \textit{support} our work practices rather than replace them.

In conclusion, software systems are designed for expert users, but in some domains these experts have very extensive tacit knowledge. Drawing on previous insights on tacit knowledge, we reported experiences with some of the issues in a domain where changing circumstances, users' `gut instincts' and extensive experience, and physically challenging environments are important issues when seeking to identify unknown knowns and unknowns.

\subsubsection{Acknowledgements.} 
We thank the participants of this research for sharing their insights, and the surveyors whom we observed for their patience and insights. We are grateful to DNV for funding this research together with The Norwegian Research Council (grant number: 309631). 
For the purpose of Open Access, the authors have applied a CC BY public copyright licence to any Author Accepted Manuscript version arising from this submission.

\end{document}